\documentclass[%
superscriptaddress,
reprint,
 amsmath,amssymb, aps,
]{revtex4-1}

\usepackage{graphicx}
\usepackage{dcolumn}
\usepackage{bm}

\begin{document}


\title{A note on quantum safe symmetric key growing}
\author{Alexander Ling}
 \email{cqtalej@nus.edu.sg}
 \affiliation{Centre for Quantum Technologies, National University of Singapore, Blk S15, 3 Science Drive 2, 117543 Singapore}
 \affiliation{Department of Physics, National University of Singapore, Blk S12, 2 Science Drive 3, 117551 Singapore}
 \author{Christoph Wildfeuer}
 \affiliation{Institute of Microelectronics, University of Applied Sciences Northwestern Switzerland, CH-5210 Windisch, Switzerland}
\author{Valerio Scarani}%
 \affiliation{Centre for Quantum Technologies, National University of Singapore, Blk S15, 3 Science Drive 2, 117543 Singapore}
 \affiliation{Department of Physics, National University of Singapore, Blk S12, 2 Science Drive 3, 117551 Singapore}

\date{\today}

\begin{abstract}
Quantum key distribution is a cryptographic primitive for the distribution of symmetric encryption keys between two parties that possess a pre-shared secret. Since the pre-shared secret is a requirement, quantum key distribution may be viewed as a key growing protocol. We note that the use of pre-shared secrets coupled with access to randomness beacons may enable key growing which, though not secure from an information-theoretic standpoint, remains quantum safe.
\end{abstract}

\maketitle



Quantum key distribution \cite{gisin02,diamanti15} (QKD) is a family of optical techniques that enable two parties possessing a pre-shared secret to grow a private symmetric key. The only additional requirement to the pre-shared secret is access to a high quality source of entropy via a quantum channel. The entropy source is either controlled by one party or, when entanglement is used, may be operated by a third-party that is not necessarily trusted. The quantum channel, typically an optical link, is not controlled.

The protocols for performing QKD are well-established and numerous trials have been conducted to demonstrate the practicability of the concept. Much effort has gone into the development of QKD to extend the range of optical links \cite{schmitt07,scheidl09,nauerth13,pugh16}, to improve the protocols for growing the secret key, and increasingly to identify and close side-channels in actual implementations \cite{lamas-linares07,nauerth09,gerhardt11}. 

The main appeal for QKD is that it provides information-theoretic security, or more simply `forward security'. If a secret key is grown correctly using QKD today, then it remains secure against any future improvements in computational power. Once in possession of this highly secure key, the users must make the decision of how to use it. The common assumption is that QKD users will deploy the key for information-theoretic encryption such as in a Vernam cipher. However, this will severely limit the quantity of data that can be encrypted, and is likely to be used only for the most critical applications \cite{etsiqkd}.

There is growing acceptance that the key grown using QKD will be used in some type of fast symmetric encryption that provides a useful level of communications performance while ensuring security against the emergence of quantum computers. This is known as ``quantum safe'' encryption. Such an encryption technique generally cannot provide full forward security as it remains vulnerable to unlimited computational power, but may be sufficient for a wide range of communications data that need to be kept private for a limited period of time.

In particular, a block-cipher scheme called the Advanced Encryption Standard (AES) has been commonly identified to be quantum safe as long as the length of the seed key is of sufficient length \cite{etsiquantumsafe,augot15}. Existing recommendations are that the seed key be doubled from 128 bits to 256 bits \cite{augot15} leaving aside the question of how the keys are distributed and managed in the first place. 

We note the possibility that methods like AES may be used to grow encryption keys in a `quantum safe' manner, without the need for any quantum communication.

\begin{figure*}[t]
	\includegraphics[width=0.9\linewidth]{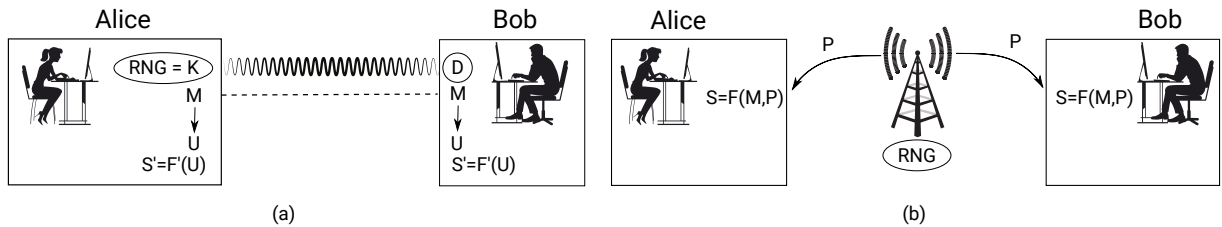}
    \caption{\label{fig:concept} Quantum safe symmetric key growing using (a) QKD or (b) a randomness beacon. In the QKD approach Alice operates a random number generator (RNG) to encode a raw key (K) into a stream of quantum particles that is transmitted to Bob's detector (D). Bob measures the particles and communicates with Alice via a conventional channel authenticated with their pre-shared secret (M) to derive a key with unconditional security (U). A cryptographic function (F') may be used to obtain an expanded session key (S'). In QKD, the pre-shared secret is used in an auxiliary protocol to prevent man-in-the-middle attacks, and its entropy is not inherited by U or S'. The RNG must be under the control of one party unless they are performing entanglement-based QKD \cite{E91}. In the beacon approach, the RNG is operated at the beacon. Random numbers in the form of pulses (P) are publicly broadcast. Alice and Bob use the pulses and their pre-shared secret (M) as inputs to a cryptographic function (F) to generate the session key (S). No quantum channel is required in this case, and multi-party key growing is possible. In both cases, a quantum safe key is obtained; in the QKD case the amount of unconditionally secure entropy inherited by the derived key S' is at best of the same length as U.} 
\end{figure*}

We consider the situation where Alice and Bob (our two parties) have a pre-shared secret, as in QKD, but instead of sharing an optical link, they now have access to a randomness beacon that both of them agree to use. The concept is illustrated in Figure~\ref{fig:concept}. A randomness beacon \cite{nistbeacon} is a broadcast service of high quality random numbers that are typically obtained from a quantum source.  Each new number that is broadcast is known as a pulse. Randomness beacons can provide pulses via the Internet or radio and may also provide an authenticated historical record of its performance to allow users to gauge the quality of the randomness being provided. 

Suppose that Alice and Bob treat their pre-shared secret random numbers as a master key (M), which they split into blocks of fixed size labelled $\textnormal{M}_\textnormal{1}$, ..., $\textnormal{M}_\textnormal{n}$. They label the pulses from the randomness beacon as $\textnormal{P}_\textnormal{1}$, ..., $\textnormal{P}_\textnormal{n}$. In a sense, Alice and Bob treat each pulse as an encrypted random number, which only they can decrypt uniquely given their pre-shared secret. These decrypted pulses can be used by Alice and Bob as session keys for normal communications. In other words, a session key $\textnormal{S}_\textnormal{i}$ would be given by $\textnormal{S}_\textnormal{i}=\textnormal{F}(\textnormal{M}_\textnormal{i},\textnormal{P}_\textnormal{i})$, where the operation F can be any quantum safe cryptographic block cipher or function. Such a method is similar to recommendations for key derivation using pseudorandom functions \cite{dworkin05} with the exception that in our example the function should be quantum safe \cite{amy16}.

As an example consider the scenario when Alice and Bob choose to use AES both for key growing and encrypting their regular communications. They need to keep in mind that while the quantum safe AES master key $\textnormal{M}_\textnormal{i}$ has to be at least 256 bits long, the function F can only process pulses $\textnormal{P}_\textnormal{i}$ that are only 128 bits long (by definition). Hence Alice and Bob will need to concatenate the output of a sequence of operations to obtain a session key $\textnormal{S}_\textnormal{i}$ that is again quantum safe. Alternative options for key growing would be to use functions derived from the Secure Hash Algorithm 3 (SHA-3) \cite{kelsey16}, or related functions.

In such a scheme, the security is entirely dependent on the choice of function F and as in any conventional encryption system, the decryption key $\textnormal{M}_\textnormal{i}$ must be changed periodically. This is not different from the case in QKD where the pre-shared secret must also be replaced periodically. Furthermore, if the output of QKD is used as session keys in a block cipher such as AES those keys must also be replaced periodically.

Another area where the decryption algorithm is important is in the possibility of collisions. There may exist scenarios where an attacker has gained control of the beacon and also knows when Alice and Bob wish to update their session key. By transmitting a pulse with a specific sequence, in the instance that Alice and Bob are updating their session key, the attacker may be able to get the decryption algorithm to produce a guessable outcome. However, this can be overcome if the public beacon maintains an authenticated database of all pulses broadcast previously, allowing Alice and Bob to choose randomly from the historical record. In this sense, the `freshness' of the pulses is irrelevant. In any case the quality of the quantum random number generator in QKD should also be subjected to scrutiny and investigated for possible device interference.

The ability of symmetric key block ciphers to resist quantum attacks is being investigated actively \cite{roetteler13,kaplan15,grassl15}. A recent paper estimated the resources needed to implement the Grover search algorithm to break AES blocks \cite{grassl15} that use key sizes of 128, 192 and 256 bits respectively. The paper provided estimates on qubit number and gate interactions needed to break AES using Grover once pre-identified ciphertext-plaintext pairs had been provided. The need for pre-identified pairs is a challenge for cryptanalysts \cite{roetteler13,grassl15}, but it should be noted that a general weakness for encryption systems occur when pairs of plaintext-ciphertext are made available (and is generally called the `known plaintext attack'). 

The use of a randomness beacon for quantum safe symmetric key growing has two major similarities to QKD (see Table 1). These are the use of pre-shared secrets between Alice and Bob, and a common source of random numbers. The difference is that a quantum channel is not required. Without the quantum channel one loses the possibility to achieve information-theoretic security. However, this key agreement method is extremely robust against environmental effects since the users can obtain the pulses via radio-frequency or internet services; high-speed data encryption can be carried out without having to rely on specialist technology; and key management techniques may enable multiple parties to derive a common session key, overcoming an existing bottleneck in QKD. 

It is interesting to consider the use of pre-shared secrets with authenticated shared entropy sources such as randomness beacons in the context of quantum safe key growing and other related cryptographic applications.
Randomness beacons that act as shared entropy sources may play a role analogous to that of the certificate authority in present day public key encryption infrastructure, and it may be worthwhile to explore the use of these beacons as part of a future quantum safe communications system.

\section*{Acknowledgements}
AL would like to thank Artur Ekert, Rene Peralta and John M. Kelsey for helpful discussions. CW would like to thank Willi Meier for valuable discussions. 

\begin{table}[h]
\caption{Beacon Key Distribution (BKD) vs. QKD.}
\begin{tabular}{l|r|r|}\hline
& BKD & QKD \\ \hline
Multi-party Keys & Y & N \\ \hline
Quantum Channel Requirement & N & Y \\ \hline
Pre-shared Secret & Y & Y \\ \hline
Shared Entropy Source & Y & Y \\ \hline
Information-Theoretic Security & N & Y \\ \hline
\end{tabular}
\end{table}

\bibliography{qskd}
\end{document}